# A Novel General Compact Model Approach for 7nm Technology Node Circuit Optimization from Device Perspective and Beyond

Qiang Huo, Zhenhua Wu, Weixing Huang, Xingsheng Wang, *Senior Member, IEEE*, Geyu Tang, Jiaxin Yao, Yongpan Liu, Feng Zhang, Ling Li, and Ming Liu, *Fellow, IEEE*

*Abstract*—This work presents a novel general compact model for 7nm technology node devices like FinFETs. As an extension of previous conventional compact model that based on some less accurate elements including one-dimensional Poisson equation for three-dimensional devices and analytical equations for short channel effects, quantum effects and other physical effects, the general compact model combining few TCAD calibrated compact models with statistical methods can eliminate the tedious physical derivations. The general compact model has the advantages of efficient extraction, high accuracy, strong scaling capability and excellent transfer capability. As a demo application, two key design knobs of FinFET and their multiple impacts on RC control ESD power clamp circuit are systematically evaluated with implementation of the newly proposed general compact model, accounting for device design, circuit performance optimization and variation control. The performance of ESD power clamp can be improved extremely. This framework is also suitable for path-finding researches on 5nm node gate-all-around devices, like nanowire (NW) FETs, nanosheet (NSH) FETs and beyond.

## I. INTRODUCTION

BSIM-CMG is the most widely used industry-standard compact model for FinFET and other ultra-scaled devices such as NW FETs and NSH FETs [1]. However, the BSIM-CMG as well as some other traditional compact models have some limits that get more serious as the device keep scaling down. First, these traditional compact models are based on one-dimensional Poisson equation, which will lead to inaccurate prediction of important parameters such as threshold voltage, capacitance and so on [2]. Second, in advanced CMOS technology node, device performance gain with scaling diminished due to undesired short channel effects, quantum effects, device variation, et al [3]-[6]. All these nonideal physical effects of sub-7nm nodes devices are difficult to be accurately integrated into the compact model directly due to the underlying physical complexity. At the same time, the model introduces a lot of parameters for these physical effects. The complex extraction procedure for devices with

This work is supported by the National Key Research Plan of China (no. 2018YFB0407500) and the National Natural Science Foundation of China under Grants 61774168, 61720106013 and 61474134.

Q. Huo, Z. Wu, W. Huang, J. Yao, F. Zhang, G. Tang, L. Li, and M. Liu are with the Key Laboratory of Microelectronics Device & Integrated Technology, Institute of Microelectronics of Chinese Academy of Sciences, Beijing 100029, China, and with the University of Chinese Academy of Sciences, Beijing 100049, China. (Corresponding author e-mails: lingli@ime.ac.cn, wuzhenhua@ime.ac.cn, zhangfeng_ime@ime.ac.cn, xswang@hust.edu.cn)

X. Wang is with Huazhong University of Science and Technology, Wuhan, China.

Y. Liu is with Tsinghua University, Beijing, China.

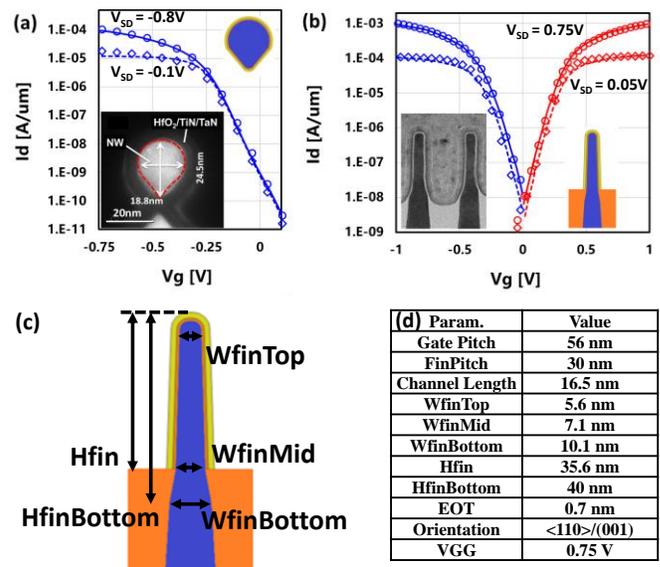

Fig.1 Model validation with reference to the linear and saturated transfer characteristics of (a) our in-house GAA Si Nanowire p-MOSFET for beyond 5nm node [19], and (b) state-of-the-art industry-standard 7nm node Si FinFET [20]. Line: Exp.; Symbol: Simu.; Inserts compare the TEM image and the corresponding simulation domain. (c) The schematic of partial parameters of FinFET. (d) Key design rules as process of record (POR) of 7nm node FinFET in this study according to [20].

different gate lengths (Lg) is provided in the BSIM-CMG manual [7], but no extracting method for different Fin widths (Wfin) is given. Last, in order to handle different ultra-scaled devices, the compact model may need a lot of modification. When a device is transferred to another device, the compact model must be physically corrected, which requires specialized experience and is difficult to maintain the same accuracy [8]. In conclusion, more universal and sophisticated compact model approach has been strongly desired for advanced technology nodes that can avoid cumbersome physical formula corrections.

Standard cell designers can change the width of a planar transistor, but they cannot change the height or width of a Fin. Channel length variation is also limited in value due to the intrinsic characteristics of the FinFET technology [9]. Although the performance of circuit can be adjusted by the number of Fins, changing the number of Fins will bring huge area overhead. It is important to further improve performancebased on the optimization of process related parameters with the area of circuit unchanged [10], especially for area-consuming circuits like SRAM and ESD power clamp. Process related parameters can be Lg, Wfin, thickness of spacer (TSPC), channel overlap or underlap length (Lov), effective oxide thickness (EOT) and so on. However, the tuning of process parameters in long range will reduce the precision of BSIM-CMG. What's worse, some device boosters like TSPC, Lov, EOT and so on cannot be directly implemented and accurately extracted by BSIM-CMG, which will lead to inaccurate circuit simulation. That means a better approach is needed.

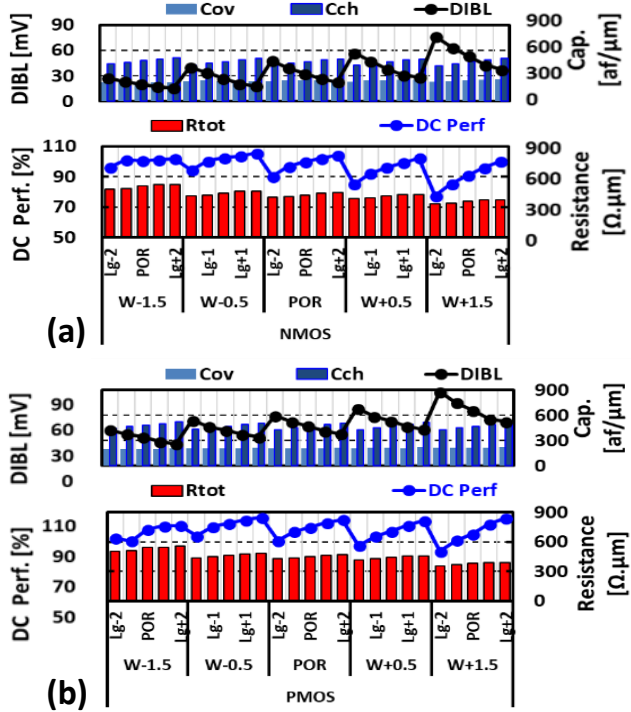

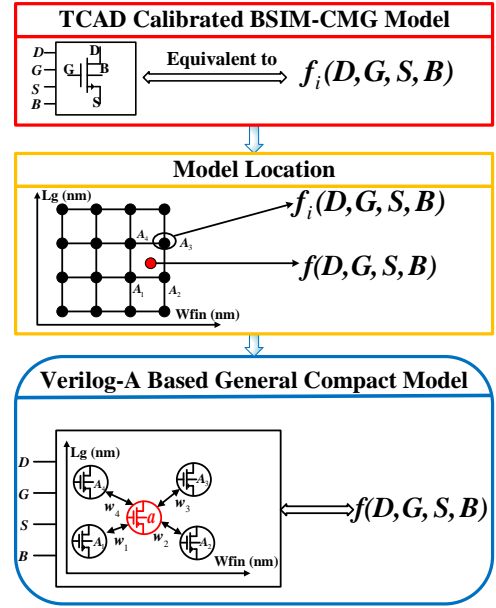

Fig.2 Key device DC and AC characteristics with knobs to the process of record (POR) design for both (a) NMOS and (b) PMOS channel shape. The changes of gate length (Lg) and Fin width (Wfin) are in unit of nm. The DC performance is characterized by the $I_{eff}$ with the target $I_{off}$.

Fig.3 A novel method of the general model integrating four TCAD calibrated BSIM-CMG models.

In this work, we propose a novel general compact model based on four TCAD calibrated compact models to predict a series of new models in a two-dimensional plane composed of Lg and Wfin. For a single device, BSIM-CMG extraction will be more accurate and easier. This model has the advantages of efficient extraction, high accuracy, strong scaling capability and excellent transfer capability. ESD power clamp is one of the most important metrics of system reliability. The circuit usually occupies a large area and is easy to work in high voltage, which makes the simulation accuracy decline [11]-[13]. We mainly use it as an application to demonstrate the effectiveness of our new model approach. In this paper, we present a framework for classical RC control ESD clamp design from device perspective with the industry standard 7nm technology node and beyond.

Our paper is organized as follows. Section II explores the optimization design space of state-of-the-art 7nm node FinFET. Section III presents the building method and evaluation of our general compact model in detail. Section IV analysis the effects of Lg and Wfin parameters on the key merits of ESD power clamp. These metrics include the clamped voltage on the VDD pad (Clamp Voltage), the quiescent VDD to GND leakage current (Leakage), the recovery time in face of false-triggering (Recovery Time) and the current drawn during power-up (Peak Power-up Leakage Current) [11]-[13]. Section V gives the conclusion.

## II. DEVICE PERFORMANCE ANALYSIS

In this work we address the quantum confinement and the ballistic transport in the FinFET on a physical level by implementing 2D Schrodinger-Poisson solver, combined with the conventional drift-diffusion equation solver for low-field region and phase-space subband Boltzmann transport equation solver for saturation region respectively [14][15]. k·p based multi-subbands electronic structures are calculated for electrostatics. The calculation of carrier scattering is based on Fermi's Golden Rule by evaluating the square matrix element of various scattering potentials and finding the corresponding transition rate. Accounting for the transition rate and screening, the energy relaxation time and the physical mobility are accurately evaluated [16]. As compared to conventional TCAD framework, the physical level modeling has advantages that it captures the subband variations under strong confinement and considers multiple carrier scattering mechanisms, i.e., phonon scattering (acoustic, optical, and intervalley models), ionized impurity scattering, and surface roughness scattering directly, rather than empirical mobility models [17][18]. Fig. 1 shows the overall verification of above physics-based models by comparing with the measurement data of the in-house 5nm node SOI nanowire [19]. Furthermore, this validated TCAD framework can well reproduce the published data of the state-of-the-art 7nm node FinFET [20] and fairly guarantee the accuracy of our general compact model and 7nm node ESD power clamp proposed in this work. The impact of 3D nature of nanoscale FinFET on circuit performance is well captured spontaneously because all our TCAD simulations are based on 3D FinFETs. Note that gate pitch and Fin pitch are fixed in our study to ensure the area invariant.

Fig. 2 summarizes the key device characteristics with design knobs of Wfin and Lg. The DC performance is characterized by the $I_{eff}$ with the target $I_{off}$ [21]. The trends of DIBL and resistance agree well with the quantum effects impacted bandgap and mobility in the narrow channel. When Fin width reduces, DC performance first increases thanks to improved DIBL and then decreases due to seriously electron mobility degradation. Such electron mobility degradation is attributed to strong quantum confinement and valley-wise carrier distributions in ultra-scaled channels [22][23], and verified by directly calculating the full-band scattering rates combined with k·p-based subband structures. Optimal effective current and capacitance at the device level can be accurately predicted.

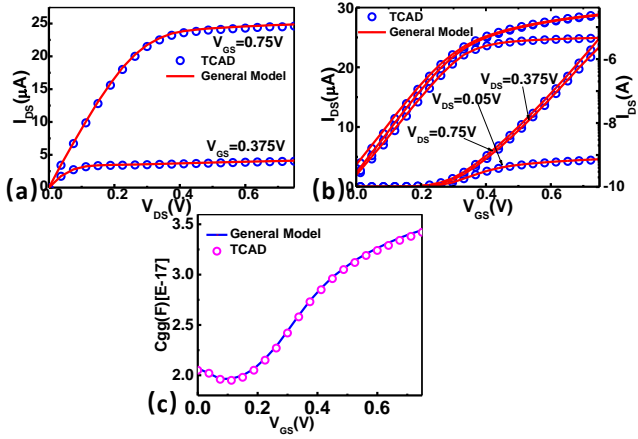

Fig.4 The validation of (a)(b) DC and (c) AC characteristics of general compact model for FinFET when Lg=17.8nm and Wfin=6.3nm. $C_{gg}$ consists of $C_{ov}$ and $C_{ch}$.

### III. GENERAL COMPACT MODEL APPROACH

In view of the shortcomings of BSIM-CMG presented by Section I, it is necessary to build hundreds of TCAD calibrated BSIM-CMG models for different Lgs and Wfins in a wide-enough-range for an accurate prognosis of circuit performance, which would be tedious, time consuming and impractical. In this section, we innovatively propose a general compact model on solving problems, which can accurately fit experimental data with the help of few decent TCAD calibrated BSIM-CMG models. This idea is inspired by ensemble learning which makes full use of the advantages of each elements and statistical methods [24].

#### A. General Compact Model Building

As shown in Fig. 3, for straightforward and easy understanding, the BSIM-CMG model can be regarded as a function $f_i(D, G, S, B)$ constructed by Verilog-A [7], where $D$, $G$, $S$ and $B$ represent drain, gate, source and bulk respectively and $i$ represents the index of BSIM-CMG model. Then, a grid is made according to the range of process parameters such as Lg and Wfin. The location of general compact model can be determined by process parameters easily. Furthermore, the general compact model $f(D, G, S, B)$ can be built by four nearest neighbor BSIM-CMG models, which can be described by $f(D, G, S, B) = \sum_{i=1}^{4} w_i f_i(D, G, S, B)$ where $w_i$ is the weight of the corresponding TCAD calibrated BSIM-CMG model $f_i(D, G, S, B)$. For calculating $w_i$ naturally, we introduce the Euler distance $dis(a, A_j)$ between general model and TCAD calibrated model as $dis(a, A_j) = [(Lg - Lg_j)^2 + (Wfin - Wfin_j)^2]^{1/2}$ where $A_j = (Lg_j, Wfin_j)$ and $a = (Lg, Wfin)$ are the position vectors of TCAD calibrated model and general model to be built respectively. It is natural to take the reciprocal of the Euler distance $dis(a, A_j)$ and then normalize it considering that the closer the distance, the greater the contribution on the general compact model. Consequently, $w_i$ can be represented as $w_i = [1/dis(a, A_i)]/\sum_{j=1}^{4}[1/dis(a, A_j)]$.

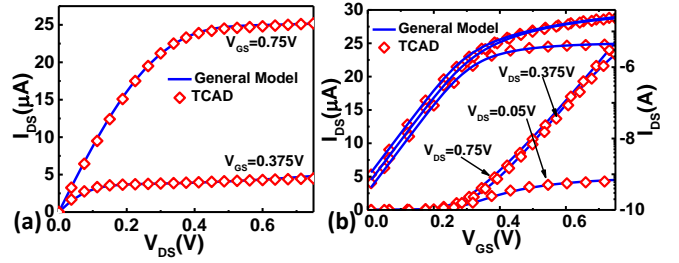

Fig.5 When the FinFET is scaled from Lg=17.8nm and Wfin=6.3nm to Lg=14.5nm and Wfin=5.1nm, general compact model still exhibits good fitting quality.

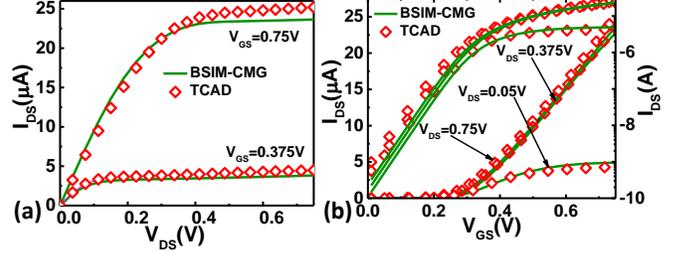

Fig.6 When the FinFET is scaled from Lg=17.8nm and Wfin=6.3nm to Lg=14.5nm and Wfin=5.1nm, BSIM-CMG exhibits worse fitting quality.

#### B. Model Evaluation

The quality of our general compact model fitting demonstrated in Fig. 4 is verified by a random case when $a = (17.8\ nm, 6.3\ nm)$ and four nearest neighbor BSIM-CMG models are $A_1 = (17.5\ nm, 6.1\ nm)$, $A_2 = (17.5\ nm, 7.1\ nm)$, $A_3 = (18.5\ nm, 7.1\ nm)$ and $A_4 = (18.5\ nm, 6.1\ nm)$ respectively. $C_{gg}$ is the total capacitance including intrinsic capacitance and parasitic capacitance. In our work, it consists of overlap capacitance ($C_{ov}$) and channel capacitance ($C_{ch}$). It should be noted that the variation range of adjacent lattice coordinates such as A$_1$ and A$_2$ in Fig.3 is suggested to be about 1 nm. Small range will increase the number of lattices, which will increase the workload of TCAD simulation. Large range will make the general compact model unable to make full use of the information of each TCAD calibrated BSIM-CMG model. Lg and Wfin in this model can be replaced by other device performance boosters such as TSPC, Lov, EOT and so on.

Despite a quadruple increase in computation load, it will not bring too much burden and time consumption to the common circuit simulation due to the high convergence speed of BSIM-CMG itself. However, the general compact model has many outstanding advantages. First, our new model is concise and efficient for extraction. This statistical model eliminates the complicated physical formulas and is very simple, easy to understand and use. More importantly, the model converts complex simultaneous extraction of multiple devices with different Lgs and Wfins into simple extraction of a single

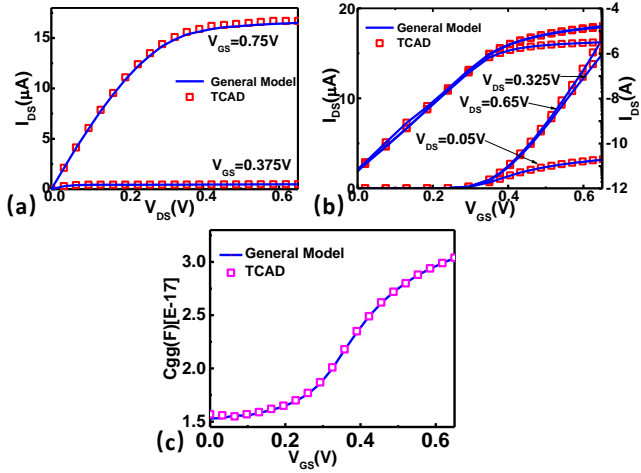

Fig.7 The validation of (a)(b) DC and (c) AC characteristics of general compact model for 5nm node nanowire FET when Lg=17.3nm and D=6.7nm. $C_{gg}$ consists of $C_{ov}$ and $C_{ch}$.

Table I. The BSIM-CMG fitting error with different Lg and Wfin.

| Lg (nm) | Wfin (nm) | I-V fitting error (RMS) |
|---|---|---|
| 18.5 | 7.1 | 2.10% |
| 17.8 | 6.3 | 0.83% |
| 16.5 | 5.6 | 1.10% |
| 14.5 | 4.1 | 2.42% |

Table II. The details of ESD power clamp performance using POR devices at 25℃.

| Performance | Value |
|---|---|
| Clamp Voltage | 0.72 V |
| Leakage Current | 3.60 μA |
| Peak Power-up Leakage Current | 55.6 mA |
| Recovery Time | 1.27 μs |

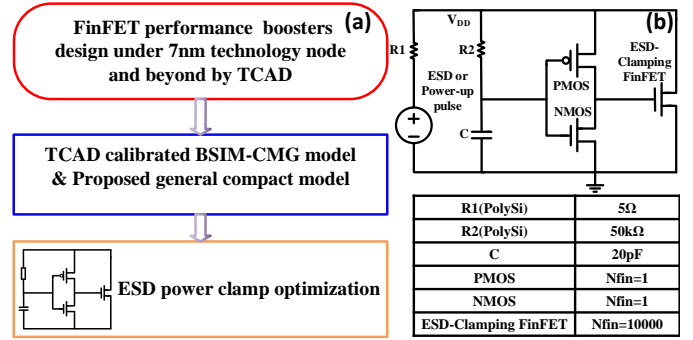

Fig.8 (a) The proposed framework for RC control ESD power clamp optimization based on our general compact model. (b) Simulation circuit and the configuration of RC control ESD power clamp.

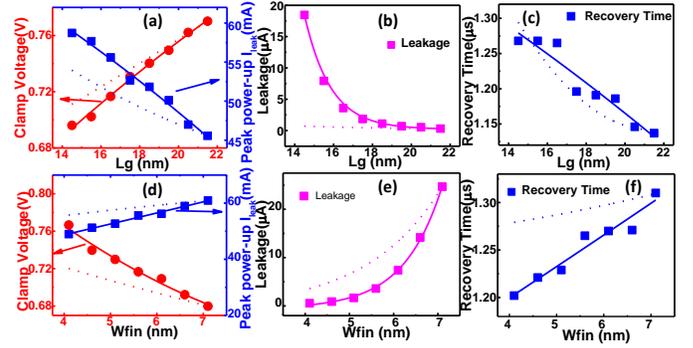

Fig.9 The impacts of (a)(b)(c) Lg and (d)(e)(f) Wfin design knobs on performances of RC control power clamp based on our general compact model (solid lines) and BSIM-CMG (dotted lines). Points mean TCAD data.

device. It greatly reduces the difficulty of BSIM-CMG extraction for TCAD data. Second, the general compact model has high accuracy and strong scaling capability. As depicted in Fig. 4, our model can accurately fit the experimental data as BSIM-CMG does for a single device. The accuracy of this model based on statistical method rather than physical formulas is determined by the neighbor TCAD calibrated BSIM-CMG models and lattice size. Since the BSIM-CMG extraction is easy and accurate for a single device, lattice size is the main determinant of accuracy. It also has strong scaling capability. When the FinFET is scaled from Lg=17.8 nm and Wfin=6.3 nm to Lg=14.5 nm and Wfin=5.1 nm, Fig. 5 demonstrates that our general compact model can still fit the experimental data very well. For comparison, the BSIM-CMG with different Lgs and Wfins is extracted as shown in Table I. This model is obtained by minimizing the relative mean square (RMS) error for FinFET with Lg=17.8 nm and Wfin=6.3 nm, while ensuring the RMS error for other devices is small and the subthreshold region fitting is as accurate as possible. However, BSIM-CMG exhibits worse scaling capability when Lg=14.5 nm and Wfin=5.1 nm in Fig. 6. The overall I-V fitting error increases to 2.88%, especially for subthreshold slope. This is mainly due to the simplification of BSIM-CMG, which makes it impossible to accurately fit the Fin-width-dependent quantum confinement. The difference in the saturation region is due to the simplification of short channel effects. As the range of Lg and Wfin increases, the BSIM-CMG extraction is more difficult and inaccurate, which will result in greater fitting error. In contrast, the general compact model still maintains good accuracy because its scaling capability depends only on the number of lattices. Last, the general compact model has good transfer capability. As mentioned earlier, the accuracy and scaling capability of our model are determined only by the size and the number of lattices and the neighbor TCAD calibrated compact models. If the compact model extraction for a single device is sufficiently accurate, our statistical-based approach can make the fitting error for any device approximate and little by tuning the size of lattices according to different processes and devices. Fortunately, the compact model extraction usually is easier and accurate enough for a single device. That means our approach can be easily transferred to other devices such as nanosheet FETs, nanowire FETs and other ultra-scaled devices while maintaining the high accuracy. Fig. 7 gives the validation of transfer capability when our model is applied to 5nm node nanowire FETs. Wfin of FinFET is replaced by nanowire diameter (D). The quality is verified when $a = (17.3\ nm, 6.7\ nm)$ and four nearest TCAD calibrated models are $A_1 = (17.2\ nm, 6.2\ nm)$, $A_2 = (17.2\ nm, 6.9\ nm)$, $A_3 = (17.9\ nm, 6.9\ nm)$ and $A_4 = (17.9\ nm, 6.2\ nm)$ respectively. In view of the conversion of 7nm node to 5nm node, the variation range of adjacent lattice coordinates is set as 0.7nm. What's more, it can be used to other device performance boosters such as TSPC, Lov, EOT and so on for circuit

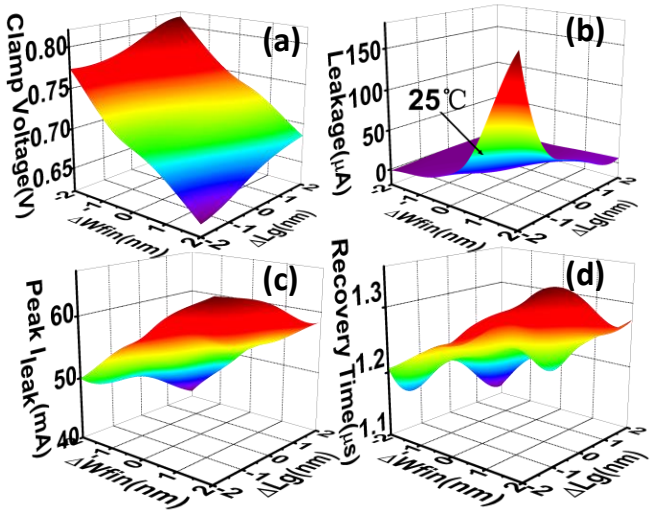

Fig.10 The impacts of FinFET cross design knobs including Lg and Wfin on RC control power clamp performance.

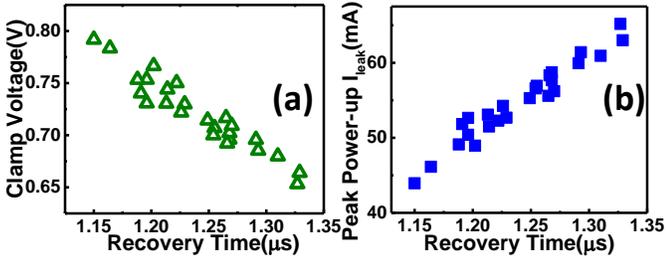

Fig.11 The correlation between (a)(b) design parameters of RC control power clamp under Lg and Wfin cross design of FinFET.

optimization from device perspective in addition to Lg and Wfin.

## IV. ESD POWER CLAMP PARAMETERS ANALYSIS

Using the BSIM-CMG model calibrated to experimental data and our proposed general compact model, we implement RC control ESD power clamp as the framework in Fig. 8(a) based on HSPICE. The simulation schematic and configuration of classical ESD clamp are shown in Fig. 8(b). The details of ESD power clamp performance using POR devices are shown in Table II for benchmarking the ESD performance. All metrics are obtained at 25℃. Similar frameworks can be used for other area-consuming circuits like SRAM with good results.

### A. General Model based Accurate Device Parameter Analysis for ESD Power Clamp

Fig.8 shows huge simulation difference between the scaling capability of BSIM-CMG including the quantum effects and short channel effects and our statistical-based general compact model on ESD power clamp. The BSIM-CMG in Fig. 9(a)(b)(c) is obtained by minimizing the relative mean square (RMS) error for FinFET with Lg=21.5 nm and Wfin=5.6 nm, while ensuring the RMS error for other devices with different Lgs is as small as possible. The BSIM-CMG in Fig. 9(d)(e)(f) extracts parameters in a similar way, focusing only on FinFET with Lg=16.5 nm and Wfin=7.1 nm. The difference between BSIM-CMG and the general compact model is further expanded at the circuit level. That further proves that although BSIM-CMG contains quantum effects and short channel effects, it is difficult to integrate precisely because of its physical complexity, which leads to the decline of scaling capability. Obviously, our model has greatly improved the accuracy of circuit simulation.

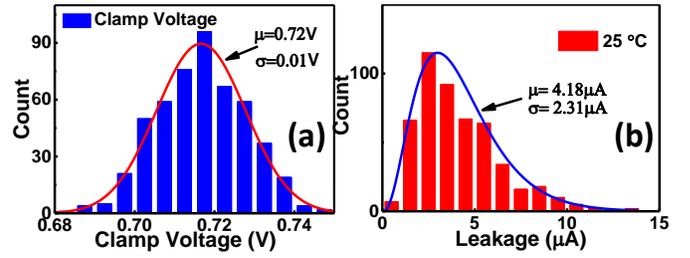

Fig.12 Statistical characteristics of (a) clamp voltage and (b) leakage current under process fluctuation condition.

As Lg decreases, the short channel effects become more significant. Recovery time increases in Fig. 9(c) is due to the ascend of time constant $R_{on}C_{gg}$. Specifically, drop in mobility caused by velocity saturation effect in the high field region (reflected in the reduction of $I_{eff}$ in DC perf of Fig. 2) exceeds the decline of capacitance $C_{gg}$ in Fig. 2. The sharp increase in leakage current in Fig. 9(b) attributes to the augment in mobility in low field region (reflected in the increase of $I_{off}$ in DC perf of Fig. 2) and drop in threshold voltage due to the increase in DIBL. The drop of clamp voltage in Fig. 9(a) is mainly due to the descend in on state resistance $R_{on}$ of the device shown in Fig. 2. With the descend of Wfin, the quantum effects become more significant and the mobility degrades obviously, but the recovery time in Fig. 9(f) decreases due to the weaker role in recovery time than the decline in capacitance $C_{gg}$. The significant drop in leakage current in Fig. 9(e) is the result of the combination of the decline in mobility and the rise in threshold voltage caused by the decrease in DIBL in Fig. 2. The increase in clamp voltage in Fig. 9(d) is mainly due to the ascend in on state resistance $R_{on}$ of the device shown in Fig. 2. Fig. 9 also depicts there are strong opposite trends between leakage current and clamp voltage, which means that the simultaneous optimization of additional power and ESD level is challenging. It is gratifying that the recovery time and the peak power-up leakage current, which play a vital function in false-triggering immunity, have strong positive correlations with leakage current, so they can be coordinated.

### B. ESD Power Clamp Optimization and Variation Control

The impacts of FinFET cross design knobs on RC control power clamp are depicted in Fig. 10, which is generated using HSPICE based on our previously proposed general compact model. The FinFET design knobs can be selected to gain the trade-off between clamp voltage, leakage current, recovery time and peak power-up leakage current according to specific needs. The relationships between design parameters are demonstrated in Fig. 11 under the joint design of FinFET Lg and Wfin. The high correlation between clamp voltage, recovery time and peak power-up leakage current means that the cooperative optimization of Wfin and Lg is easy to ensure good results. This is because all these performances are closely related to the variation of mobility. Clamp voltage and recovery time can be improved respectively by up to 8.8% and 9.4% through the trade-off between Lg and Wfin. Leakage current and peak power-up leakage current can be reduced respectively by up to 93.3% and 24.3% through the trade-off between Lg and Wfin compared to the RC control power clamp under POR condition.

Statistical characteristics of leakage current and clamp voltage under process fluctuation are illustrated in Fig. 12. We apply the empirical stochastic fluctuations of Lg and Fin width of POR device with Gauss distribution as process fluctuations [25]. Clamp voltage obeys normal distribution with minimal standard deviation basically, that is, clamp voltage which determines ESD level is immune to process fluctuation. However, due to the huge variation of mobility caused by quantum effects under process fluctuation at sub-7nm nodes, leakage current is more sensitive and tends to deteriorate. Therefore, process fluctuation is more likely to cause huge additional power dissipation while maintaining relatively steady ESD level. For these problems, we give process solutions such as H plasma annealing process to control Fin width and sidewall inclination while improving surface roughness and reducing multiple exposures to single exposure for EUV to mitigate fluctuations. These methods can effectively reduce current fluctuation and power consumption.

## V. CONCLUSION

In this study, we present a general compact model combining few TCAD calibrated compact models with statistical methods which can skip the tedious physical derivation. The model has the advantages of efficient extraction, high accuracy, strong scaling capability and excellent transfer capability. Our model can greatly improve the accuracy of circuit simulation under the joint design of different device performance boosters. As an application, we present a framework for classical RC control ESD clamp design from device perspective with the industry standard 7nm technology node and beyond. It is found that the trade-off between Lg and Wfin of FinFET can greatly strengthen the performance of power clamp while remaining the area unchanged, and leakage current is seriously affected and tends to deteriorate under process fluctuation. We firmly prove that the framework based on our general compact model are effective for accurate circuit optimization under state-of-the-art technology node and beyond.